\documentclass[twocolumn,nofootinbib,amsmath,amssymb,a4paper]{revtex4}

\usepackage{graphicx}
\usepackage{dcolumn}
\usepackage{bm}

\begin{document}

\title{Lattice calculations for ${\mathbf B}$ and ${\mathbf K}$ mixing}

\author{N. Tantalo}
\email{nazario.tantalo@roma2.infn.it}
\affiliation{%
INFN sezione di Roma ``Tor Vergata'', I-00133 Rome, Italy\\
Centro Enrico Fermi, Via Panisperna 89 A, I-00184 Rome, Italy
}%

\begin{abstract}
The bag parameters and the decay constants of neutral $B_{(s)}$ and $K$ mesons were among the 
non-perturbative hadronic inputs to the classical CKM Unitarity Triangle Analysis. Thanks 
to the big amount of experimental information collected in the last few years at the 
$B$-factories and by the CDF collaboration, these matrix elements are now among the outputs of
the unitarity fits, once the validity of the Standard Model has been postulated. 
Lattice calculations of the mixing amplitudes are still needed in order to make a test of the 
theory, provided that their statistical and systematic errors are under control at the level 
of a few percent. Here we review some of the recent lattice calculations of these quantities.
\end{abstract}

\maketitle

%ooooooooooooooooooooooooooooooooooooooooooooooooooooooooooooooooooooooooooooooooooooooooooooo
\section{Introduction}
Lattice QCD calculations of the $B_{(s)}$ and $K$ mixing amplitudes where needed in the past
in order to be used as inputs to the Unitarity Triangle Analysis (UTA).
After the recent measurements of the angles of the unitarity triangle
the $\rho$-$\eta$ plane is over-constrained and the mixing amplitudes can now be extracted by 
one of the possible variants of the UTA (see~\cite{Bona:2006ah} and references there).
From the phenomenological point of view it is thus legitimate to ask whether 
it is still needed a precise lattice calculation of the mixing amplitudes.
In answering this question the point to note is that the unitarity fits are 
over-constrained only within the Standard Model. In this scenario one can make a precision 
test of the Standard Model and hopefully reveal the presence of new physics by comparing 
the  ``experimental'' determinations of the mixing amplitudes with their theoretical 
predictions. To this end lattice is still needed but it has to provide, let's say, 
$\Delta M_s$ with an error of the same order of magnitude of the one quoted  by the CDF 
collaboration, i.e. all the systematic errors have to be under control at a level of a few 
percent. Among all the systematics that presently affect lattice calculations\footnote{see 
C.~Sachrajda and T.~Onogi talks at this workshop for an extended discussion on the lattice 
systematics} the worst are the  ``uncontrolled'' ones, in the sense that cannot be reliably 
quantified without performing independent calculations with different discretizations of the 
continuum action. To this category certainly belong quenching and, in our opinion,
rooted staggered fermions. Staggered fermions are introduced on the lattice by simulating 
a quark action that suffers from doubling, i.e. it has 16 one-component fermions 
in the classical continuum limit that are packaged into 4 tastes of 4-component 
Dirac fermions. Rooting means that gauge configurations are generated by taking the fourth
root of the staggered quark determinant\footnote{This formalism has been largely used because
it is particularly cheap from the computational point of view but recently it has been
proved the feasibility of large scale $N_f=2$ lattice simulations with pions as light
as 300~MeV and physical volumes of the order of 2~fm.}.
If ``taste symmetry'' is exact, like in the continuum theory, rooting is obviously a kosher 
operation. At finite lattice spacing taste symmetry is broken and rooting breaks locality 
and unitarity. Whether these properties are recovered or not by taking the continuum limit
is the subject of a strong debate within the lattice community. 
At the annual lattice conference S.~Sharpe has given a beautiful critical
review on the subject~\cite{Sharpe:2006re} entitled
``Rooted staggered fermions: Good, bad or ugly?''. His answer to this
question is ``ugly, in the sense that they are affected by unphysical contributions at 
regulated stage that need a complicate analysis to be removed''. We share his view.

%ooooooooooooooooooooooooooooooooooooooooooooooooooooooooooooooooooooooooooooooooooooooooooooo
\section{$\mathbf B_K$}
The so-called bag parameter $B_K$ parametrizes the mixing amplitude of the $\bar{K}_0$-$K_0$
mesons according to 
\begin{eqnarray}
  &&\langle \bar K^0 \vert \hat O_1 \vert K^0 \rangle =
  \langle \bar K^0 \vert \hat O_{VV+AA} \vert K^0 \rangle  = 
	  {8\over 3}M_{K}^2  f_{K}^2\; B_{K}(\mu)
  \nonumber
\end{eqnarray}
where the operator $O_1=\bar s^i \gamma_\mu (1- \gamma_{5} )  d^i\;
\bar s^j  \gamma_\mu (1- \gamma_{5} ) d^j$ is usually conveniently decomposed into a 
parity-even and a parity-odd part, $O_1=O_{VV+AA}-O_{VA+AV}$. 

The breaking of chiral symmetry on the lattice with Wilson fermions complicates 
considerably the renormalization pattern of the $O_{VV+AA}$ operator that happens to mix 
with the other 4 operators in the parity-even basis~\cite{Martinelli:1983ac,Bochicchio:1985xa,
Donini:1999sf}. The issue in such a calculation consists in obtaining the numerical accuracy 
required in order to keep under control the mixing subtractions or in devising smart 
strategies to avoid the problem.
Two groups have been able to circumvent the mixing problem. The authors of 
refs.~\cite{Guagnelli:2005zc,Dimopoulos:2006dm} have used the so called twisted mass
formulation of lattice QCD (tmQCD) in order to map the matrix element of the parity even 
operator $O_{VV+AA}$ into the matrix element of the
parity odd operator $O_{VA+AV}$ that renormalizes multiplicatively; they have been able
to calculate $B_K$ by keeping under control all the sources of errors apart from quenching
(non-perturbative renormalization, estimate of $SU(3)$-breaking effects, continuum
limit with 5 lattice spacings, estimate of finite volume effects).
The authors of refs.~\cite{Becirevic:2002mm,Mescia:2005ew} avoided the
mixing by using a chiral ward identity that again relates the matrix elements of $O_{VV+AA}$
to that of $O_{VA+AV}$ at the price of computing on the lattice a four-point Green function.

In the case of lattice discretizations that satisfy the so called Ginsparg-Wilson (GW) 
relation an exact chiral symmetry is preserved also at finite
lattice spacing. Domain wall fermions satisfy the GW in the limit of an infinite fifth
dimension. Practically, the fifth dimension is finite and the lattice chiral symmetry is only
approximately preserved. The authors of refs.~\cite{Izubuchi:2003rp,Cohen:2006xi} have
performed a calculation of $B_K$ with respectively $N_f=2$ and $N_f=2+1$ flavours of
dynamical domain wall fermions. The $N_f=2+1$ results have been obtained at fixed lattice
spacing ($a\simeq 0.12$~fm), with non perturbative renormalization (by
neglecting the small mixing due to the ``residual mass term''), by interpolating the physical
$K$ meson state, on a single volume ($L\simeq 2$~fm); a simulation at the same lattice spacing
on a larger volume ($L\simeq 3$~fm) is under way.

\begin{figure}
\includegraphics[width=0.4\textwidth]{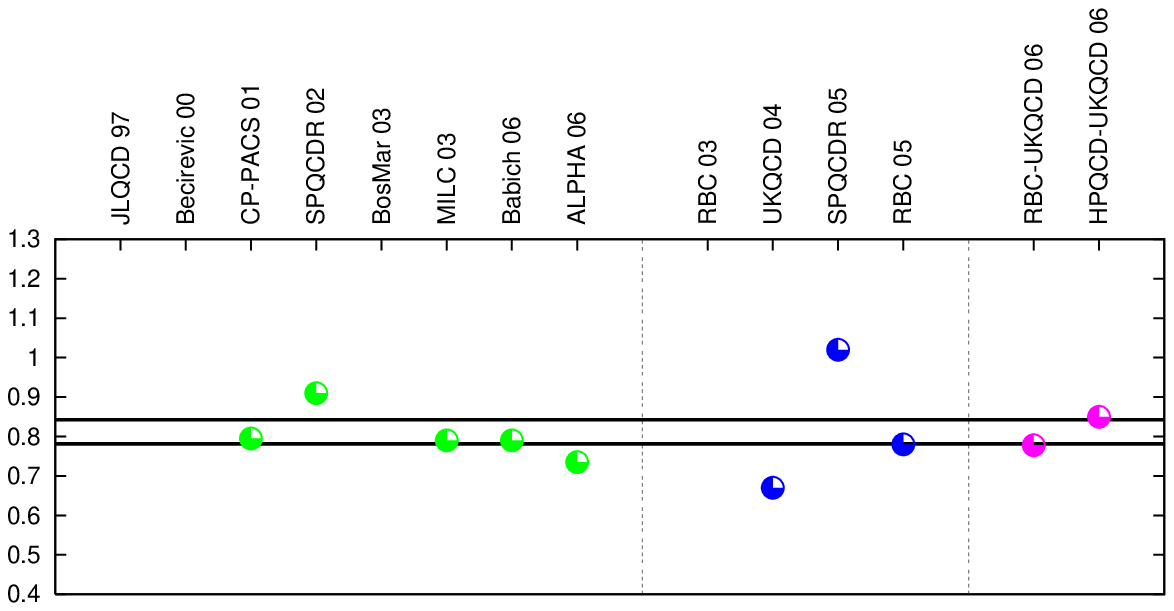}
\includegraphics[width=0.4\textwidth]{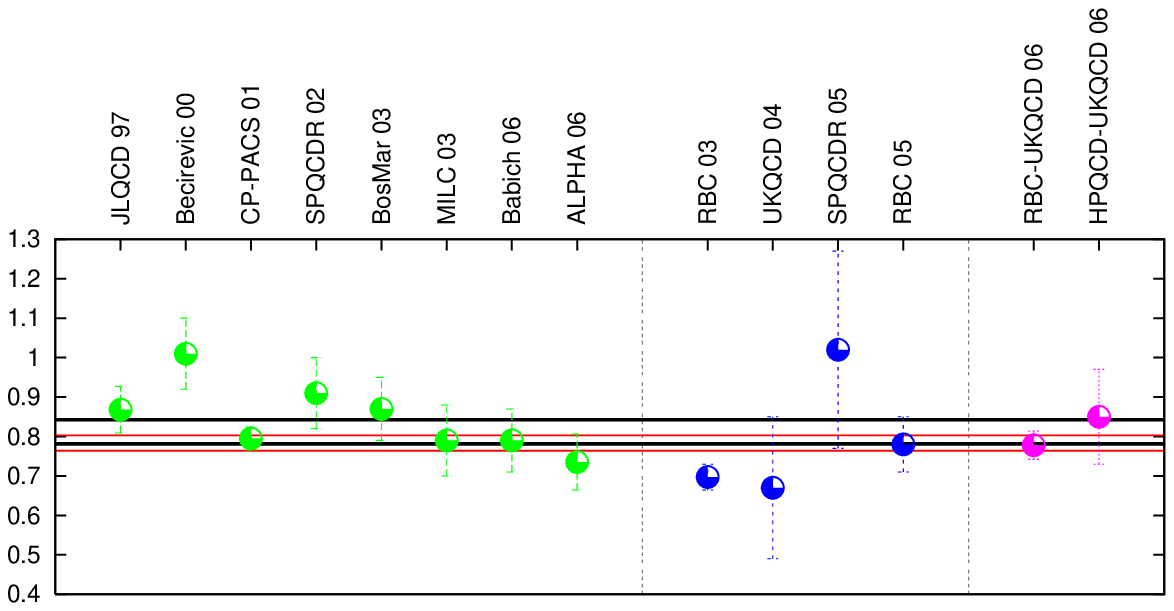}
\caption{\label{fig:BKnoerrors} 
\emph{Top}. Average of the uncorrelated $\hat{B}_K$ results without the quoted errors.
\emph{Bottom}. Average of the $\hat{B}_K$ results with $N_f>0$ with the quoted errors 
(red lines).
}
\end{figure}
\begin{table}
\caption{\label{tab:bksummary}
Lattice calculations of the renormalization group invariant (RGI) kaon bag 
parameter $\hat{B}_K$.}
\begin{ruledtabular}
\begin{tabular}{lrr}
collaboration & $\hat{B}_K$ & $N_f$ \\
\hline
JLQCD97~\cite{Aoki:1997nr}             &0.868(59)     &0\\ 
Becirevic00~\cite{Becirevic:2000ki}    &1.01(9)       &0\\
CP-PACS01~\cite{AliKhan:2001wr}        &0.795(29)     &0\\
SPQCDR02~\cite{Becirevic:2002mm}       &0.91(9)       &0\\
BosMar03~\cite{Garron:2003cb}          &0.87(8)       &0\\
MILC03~\cite{DeGrand:2003in}           &0.79(9)       &0\\
Babich06~\cite{Babich:2006bh}          &0.79(8)       &0\\
ALPHA06~\cite{Dimopoulos:2006dm}       &0.735(71)     &0\\
\hline
RBC03~\cite{Izubuchi:2003rp}           &0.697(33)     &2\\
UKQCD04~\cite{Flynn:2004au}            &0.67(18)      &2\\
SPQCDR05~\cite{Mescia:2005ew}          &1.02(25)      &2\\
RBC05~\cite{Aoki:2005ga}               &0.78(7)       &2\\
\hline
RBC-UKQCD06~\cite{Cohen:2006xi}        &0.778(36)     &2+1\\
HPQCD-UKQCD06~\cite{Gamiz:2006sq}      &0.85(12)      &2+1\\
\end{tabular}
\end{ruledtabular}
\end{table}

On the one hand, there have been so many different calculations of $B_K$ among the years
that it is not possible to enter into the details of all of them in this short 
review\footnote{we have just mentioned some representative calculations
and apologize with the authors whose results have not been covered in greater
detail. The same holds also for the following sections.} (see TABLE~\ref{tab:bksummary}). 
On the other hand none of this calculations is able to take under control all the sources of 
systematics at the same time. Since different numbers have been obtained with different
actions, techniques, assumptions, etc. we can get an estimate of the systematics by averaging 
all the results that are ``uncorrelated''
(in the sense that we neglect results that have been updated by the same collaboration
at fixed $N_f$) without the quoted errors (see FIG.~\ref{fig:BKnoerrors} top plot).
As a result we get $\hat{B}_K= 0.81(3)$ i.e. a relative error of the order of $4\%$; 
if instead we take the average of the numbers with $N_f>0$ by trusting the quoted errors 
we get $\hat{B}_K=0.78(2)$ (see FIG.~\ref{fig:BKnoerrors} bottom plot).
The previous numbers have to be taken as ``provocative'' averages: unless a
clear statement is made on which lattice results can be trusted and which have to be excluded
from phenomenological analysis one should conclude that $\hat{B}_K$ is presently predicted
by the lattice with a few percent error. A conservative estimate of the errors, to
be used in phenomenological applications, can be obtained for example by accounting for the 
dispersion of the results:
\begin{equation}
\hat{B}_K= 0.78(2)(9)
\end{equation}

%ooooooooooooooooooooooooooooooooooooooooooooooooooooooooooooooooooooooooooooooooooooooooooooo
\section{$\mathbf f_{B_q}$}
The decay constants of the $B_q$ mesons, where $q$ stays for either a down or a strange
quark, enter in the parametrization of the $\bar{B}_q$-$B_q$ mixing amplitudes together
with the bag parameters,
\begin{displaymath}
  \langle \bar B_q \vert \hat O_{VV+AA} \vert B_q \rangle  = 
	  {8\over 3}M_{B_q}^2  f_{B_q}^2\; B_{B_q}(\mu)
\end{displaymath}
What it is actually needed in order to perform 
the UTA is the combination $f_{B_q}\sqrt{B_{B_q}}$, that comes out to have a smaller 
statistical error on the lattice w.r.t. the product of $f_{B_q}$ and $\sqrt{B_{B_q}}$
computed separately. Since there are many more calculations of the decay constants than the
bag parameters and since we want to use as much information as possible in taking
the averages, we will discuss separately the lattice calculation of $f_{B_q}$, in this 
section, and of $B_{B_q}$, in the next section, but first we briefly comment on the issues
related to the simulation of the heavy flavours on the lattice. 

On currently affordable lattice sizes (at least in unquenched simulations) one has 
$am_b>1$~and~$Lm_q>1$ or $am_b<1$~and~$Lm_q<1$, i.e. a relativistic beauty-light meson
can be simulated on big volumes with big cutoff effect or on small volumes with big
finite volume effects. The different approaches that have been devised in order to 
solve this problem can be divided into ``big volume'' and ``small volume'' strategies.

\emph{Big volume strategies}. Lattice simulations of heavy quarks on physical volumes
are performed by recurring to effective field theories. 
One possibility is to simulate the lattice static action and, eventually, 
the relativistic theory with heavy quark masses in the charm region in order to interpolate
the bottom region.
Another possibility is the so called Fermilab approach~\cite{El-Khadra:1996mp,Aoki:2001ra,
Christ:2006us} that consists in improving the heavy quark action with mass dependent 
coefficients, provided that $\vert a\vec{p}\vert\ll 1$; the procedure smoothly interpolates 
between heavy and light quarks and the continuum limit can be taken although the mass 
dependence of the improving coefficients makes the procedure highly non trivial.
Still another possibility is to simulate on the lattice a discretized form of the non 
relativistic heavy quark action expanded to a given order in $v^2$ and $\alpha_s$ 
(see for example~\cite{Gray:2005ur}); as a consequence of the non-renormalizability of the
theory, the continuum limit cannot be taken and the matching with full QCD can be done only 
perturbatively; furthermore the expansion is expected to work for onium systems
but it has been widely used also to study heavy--light mesons.

\emph{Small volume strategies}. The step scaling method (SSM) has been proposed 
in~\cite{Guagnelli:2002jd} in order to deal with two-scale, $E_l\ll E_h$, problems in lattice
QCD. The method starts from the following identity
\begin{eqnarray}
  &&\mathcal{O}(E_h,E_l,\infty) =
  \nonumber\\
  &&\mathcal{O}(E_h,E_l,L_0)\times
  \underbrace{\frac{\mathcal{O}(E_h,E_l,2L_0)}{\mathcal{O}(E_h,E_l,L_0)}}_{
    \sigma(E_h,E_l,L_0)}\times
  \underbrace{\frac{\mathcal{O}(E_h,E_l,4L_0)}{\mathcal{O}(E_h,E_l,2L_0)}}_{
    \sigma(E_h,E_l,2L_0)}\dots 
  \nonumber
\end{eqnarray}
The idea is to start the computation on a small volume where the high energy scale can be set 
to its physical value and to correct for the finite volume effects by computing the step 
scaling functions, $\sigma(E_h,E_l,L)$, stopping the recursion when the last factor is equal 
to one within the required precision. The strength of the method can be better understood by 
specializing the previous identity to the case of a heavy-light meson observable, say $f_B$. 
In this case the dependence of the step scaling functions upon the $b$ quark mass can be 
predicted by using HQET both at numerator and denominator
\begin{eqnarray}
  &&\sigma(m_h,m_d,L) = \frac{f_B^0(m_d,2L)\left(1+\frac{f_B^1(m_d,2L)}{m_h}+\dots\right)}
                             {f_B^0(m_d,L)\left(1+\frac{f_B^1(m_d,L)}{m_h}+\dots\right)}
  \nonumber \\
  &&= \sigma^{\mbox{stat}}(m_d,L)\left(
  1+\frac{f_B^1(m_d,2L)-f_B^1(m_d,L)}{m_h}+\dots
  \right) \nonumber
\end{eqnarray}
The maximum number of points of the lattice discretization is fixed during the
recursion so that, in order to have $am_h\ll 1$ at each step, the step scaling functions
have to be computed at heavy quark masses smaller than the physical beauty mass.
Nevertheless, from the previous equation, it is clear that the step scaling functions
depend mildly upon the high energy scale thank to the cancellation 
$f_B^1(m_d,2L)-f_B^1(m_d,L)$ that gets stronger and stronger as the volume is increased.
Another useful application of finite volume techniques has been used, originally
in ref.~\cite{Heitger:2003nj} to implement the renormalization procedure of HQET fully 
non-perturbatively. The idea is to match the effective theory with full QCD on a small volume
at the scale $\mu=m_b$. Then the running of the matching factors is computed through a step 
scaling recursion in the effective theory. This method can be applied
in conjunction with the SSM to compute, for example, $\sigma^{\mbox{stat}}(m_d,L)$ 
(see ref.~\cite{Guazzini:2006bn}).

\begin{figure}
\includegraphics[width=0.4\textwidth]{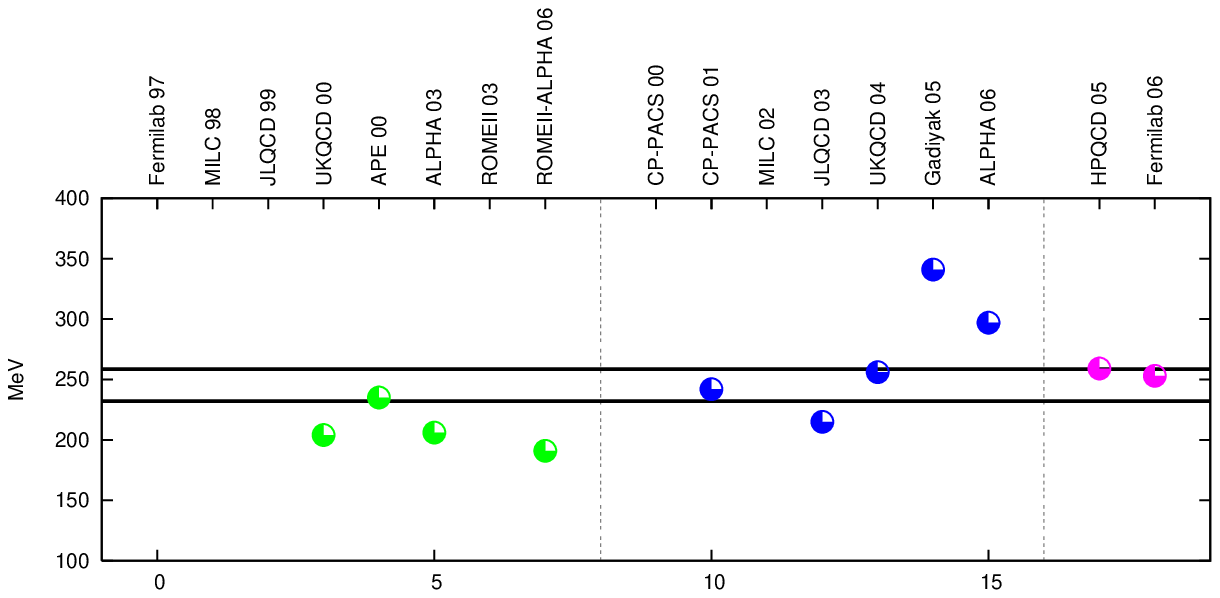}
\includegraphics[width=0.4\textwidth]{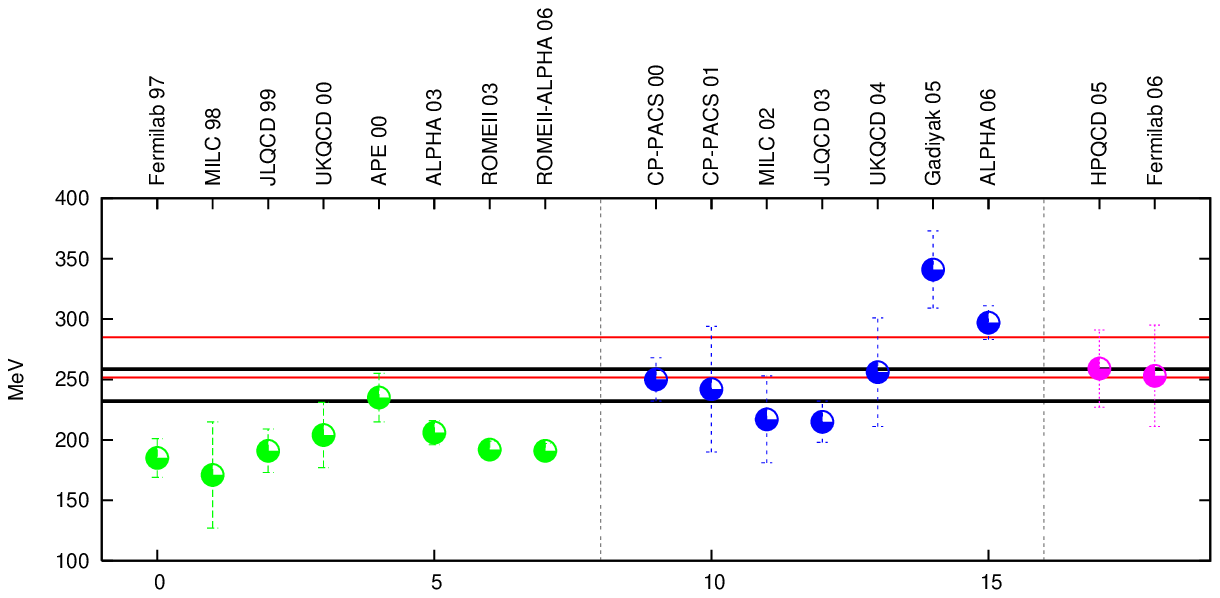}
\caption{\label{fig:fBsnoerrors} 
\emph{Top}. Average of the uncorrelated $f_{B_s}$ results without the quoted errors.
\emph{Bottom}. Average of the $f_{B_s}$ results with $N_f>0$ with the quoted errors 
(red lines).
}
\end{figure}
\begin{table}
\caption{\label{tab:fBsummary}
Lattice calculations of $f_{B_s}$ (MeV) and $r_B=f_{B_s}/f_B$.}
\begin{ruledtabular}
\begin{tabular}{lrrr}
collaboration & $f_{B_s}$ &  $r_B$ & $N_f$ \\
\hline
Fermilab97~\cite{El-Khadra:1997hq}            &185(16)  &                    &0\\  
MILC98~\cite{Bernard:1998xi}                  &171(44)  &                    &0\\  
JLQCD99~\cite{Ishikawa:1999xu}                &191(18)  &                    &0\\  
UKQCD00~\cite{Lellouch:2000tw}                &204(27)  &                    &0\\  
APE00~\cite{Becirevic:2000nv}                 &235(20)  &                    &0\\  
ALPHA03~\cite{Rolf:2003mn}                    &206(10)  &                    &0\\  
ROMEII03~\cite{deDivitiis:2003wy}             &192(7)   &                    &0\\  
ROMEII-ALPHA06~\cite{Guazzini:2006bn}         &191(6)   &                    &0\\  
\hline
CP-PACS00~\cite{AliKhan:2000eg}               &250(18)  &1.203(64)           &2\\  
CP-PACS01~\cite{AliKhan:2001jg}               &242(52)  &1.179(29)           &2\\  
MILC02~\cite{Bernard:2002pc}                  &217(36)  &1.16(5)             &2\\  
JLQCD03~\cite{Aoki:2003xb}                    &215(17)  &1.13(12)            &2\\  
UKQCD04~\cite{McNeile:2004wn}                 &256(45)  &1.38(15)            &2\\  
Gadiyak05~\cite{Gadiyak:2005ea}               &341(32)  &1.38(15)            &2\\  
ALPHA06~\cite{DellaMorte:2006sv}              &297(14)  &                    &2\\  
\hline
HPQCD05~\cite{Gray:2005ad}                    &259(32)  &                    &2+1\\   
Fermilab-MILC-HPQCD06~\cite{Simone:2006lat}   &253(42)  &1.27(6)             &2+1\\
\end{tabular}
\end{ruledtabular}
\end{table}

We now come to the lattice results for $f_{B_s}$ and $r_B=f_{B_s}/f_B$. Among the quenched
results those of refs.~\cite{deDivitiis:2003wy,Guazzini:2006bn} have been obtained through 
the SSM by keeping under control all the systematics apart from the quenching. 
Among the $N_f=2$ results, the ones of refs.~\cite{McNeile:2004wn,Gadiyak:2005ea} have been 
obtained within the static approximation giving a particularly high value of $r_B$; also the 
result of~\cite{DellaMorte:2006sv} is static but non-perturbatively renormalized. The $N_f=3$ 
results~\cite{Gray:2005ad,Simone:2006lat} have been obtained by using rooted staggered 
fermions for the dynamical light quarks and NRQCD and Fermilab respectively for the heavy. 
By looking at FIG~\ref{fig:fBsnoerrors} it emerges that quenched lattice calculations, 
though compatible within themselves, are systematically smaller than unquenched results; the 
quoted errors are still large but unquenching seems to have a significant effect on this 
observable (this is not the case of $B_K$ within the quoted errors) and the ``provocative'' 
average ($f_{B_s}= 245(13)$~MeV and $r_B=1.24(4)$) prefers unquenched results. 
The unquenched average, static points included, with an error that takes into account the 
spread of the results is
\begin{equation}
f_{B_s}= 268(17)(20)\mbox{ MeV,} \qquad
\frac{f_{B_s}}{f_{B}}=1.20(2)(5) 
\end{equation}

%ooooooooooooooooooooooooooooooooooooooooooooooooooooooooooooooooooooooooooooooooooooooooooooo
\section{$\mathbf B_{B_q}$}
In order to calculate the bag parameters of the $B_q$ mesons one has to face at the same time
the problem of the mixing, as for $B_K$, and the problem related to the presence of a heavy
and a light quark, as for $f_{B_q}$. For this reason the number of lattice calculations of
$B_{B_q}$ is much smaller than in the case of $f_{B_q}$ or $B_K$.

\begin{table}[t]
\caption{\label{tab:BBsummary}
Lattice calculations of $B_{B_s}(m_b)$ and $B_B(m_b)$.}
\begin{ruledtabular}
\begin{tabular}{lrrr}
collaboration & $B_{B_s}(m_b)$ &  $B_B(m_b)$ & $N_f$ \\
\hline
UKQCD00~\cite{Lellouch:2000tw}                 &0.90(4)    &0.91(6)          &0\\
APE00~\cite{Becirevic:2000nv}                  &0.92(7)    &0.93(10)         &0\\
SPQCDR01~\cite{Becirevic:2001xt}               &0.87(5)    &0.87(6)          &0\\
JLQCD02~\cite{Aoki:2002bh}                     &0.86(5)    &0.84(6)          &0\\
\hline
JLQCD03~\cite{Aoki:2003xb}                     &0.850(64)  &0.836(68)        &2\\
Gadiyak05~\cite{Gadiyak:2005ea}                &0.864(76)  &0.812(82)        &2\\
\hline
HPQCD06~\cite{Dalgic:2006gp}                   &0.76(11)   &                 &3\\
\end{tabular}
\end{ruledtabular}
\end{table}

Nevertheless, by looking at TABLE~\ref{tab:BBsummary} it emerges that, 
within the quoted errors, $B_{B_q}$ does not seem to depend upon the number of dynamical 
flavours, the renormalization systematics (the quenched result of ref.~\cite{Becirevic:2001xt}
has been non-perturbatively renormalized), the strategy used to handle with heavy quarks and 
even the light quark mass. 
Actually, the matrix element $\langle \bar B_q \vert \hat O_{VV+AA} \vert B_q \rangle$ does 
show a sizable dependence upon all these variables but through the vacuum saturation 
approximation, i.e. $8M_{B_q}^2 f_{B_q}^2/3$. The average of the $N_f>0$ calculations, with 
an error that takes into account the spread of the results, is
\begin{equation}
B_{B_s}(m_b)= 0.84(3)(5)\qquad
B_{B}(m_b)  = 0.83(1)(6)
\end{equation}

%ooooooooooooooooooooooooooooooooooooooooooooooooooooooooooooooooooooooooooooooooooooooooooooo
\section{a calculation of $\mathbf G(\omega)$}
\begin{figure}[h]
\includegraphics[width=0.4\textwidth]{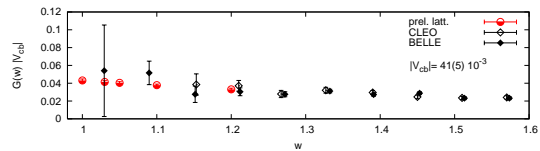}
\caption{\label{fig:VCB}
Comparison of lattice data and experimental determinations of $G(\omega)$:
$V_{cb}=41(5)\times 10^{-3}$ is extracted by using the experimental points
at $\omega=1.2$.
}
\end{figure}
We now change subject to put up to the results of a preliminary quenched 
calculation~\cite{deDivitiis:2007prep} of the form factor $G(\omega)$, needed in order to 
extract $V_{cb}$ from the exclusive semileptonic decay $B_{(s)}\rightarrow D_{(s)}\ell\nu$
 ($\omega=p_B\cdot p_D/M_BM_D$). The calculation has been carried on by using the SSM and 
by defining the form factor and the kinematical factors in terms of ratios of three point 
correlation functions. 
The results have been obtained with a relative error of about $4$\% for values of $1\le 
\omega \le 1.2$ were experimental data do not need to be extrapolated (FIG.~\ref{fig:VCB}). 

%ooooooooooooooooooooooooooooooooooooooooooooooooooooooooooooooooooooooooooooooooooooooooooooo
\begin{acknowledgments}
I would like to thank my collaborators G.M.~de~Divitiis and R.~Petronzio and also V.~Lubicz, 
F.~Mescia and T.~Onogi for interesting discussions on the topics covered in this talk. 
A special thank goes to the  organizers of the CKM~2006 workshop for the kind hospitality in 
Nagoya.
\end{acknowledgments}

%ooooooooooooooooooooooooooooooooooooooooooooooooooooooooooooooooooooooooooooooooooooooooooooo

%ooooooooooooooooooooooooooooooooooooooooooooooooooooooooooooooooooooooooooooooooooooooooooooo
\end{document}